\documentclass[10pt]{article}

\usepackage{latexsym,amssymb,amsmath,theorem}

\newlength{\dinwidth}
\newlength{\dinmargin}
\def\endexem{\hfill{$\Box$}\medskip}

\theoremstyle{change}
\theoremheaderfont{\scshape}
\newtheorem{theorem}{Theorem}[section]
\newtheorem{prop}[theorem]{Proposition}
\newtheorem{lem}[theorem]{Lemma}
\newtheorem{corollary}[theorem]{Corollary}
\newtheorem{conj}[theorem]{Conjecture}
\newtheorem{definition}[theorem]{Definition}

\newtheorem{defprop}[theorem]{Definition/Proposition}
\theorembodyfont{\rmfamily}
\newtheorem{example}[theorem]{Example}
\newtheorem{remark}[theorem]{Remark}

\newcommand{\be}{\begin{equation}}
\newcommand{\ee}{\end{equation}}
\newcommand{\bea}{\begin{eqnarray}}
\newcommand{\eea}{\end{eqnarray}}
\newcommand{\bean}{\begin{eqnarray*}}
\newcommand{\eean}{\end{eqnarray*}}

\newcommand{\bdefin}{\begin{defin}}
\newcommand{\blemma}{\begin{lem}}
\newcommand{\bprop}{\begin{prop}}
\newcommand{\btheor}{\begin{theorem}}
\newcommand{\bcoro}{\begin{cor}}
\newcommand{\bconj}{\begin{conj}}
\newcommand{\bdefprop}{\begin{defprop}}
\newcommand{\bexam}{\begin{example}}
\newcommand{\edefin}{\end{defin}}
\newcommand{\elemma}{\end{lem}}
\newcommand{\eprop}{\end{prop}}
\newcommand{\etheor}{\end{theorem}}
\newcommand{\ecoro}{\end{cor}}
\newcommand{\econj}{\end{conj}}
\newcommand{\brem}{\begin{rema}}
\newcommand{\erem}{\end{rema}}
\newcommand{\edefprop}{\end{defprop}}
\newcommand{\eexam}{\endexem\end{example}}

\def\1#1{{\bf #1}}
\def\2#1{{\cal #1}}
\def\3#1{{\sl #1}}
\def\4#1{{\tt #1}}
\def\5#1{{\sf #1}}
\def\6#1{{\mathfrak #1}}
\def\7#1{{\mathbb #1}}

\numberwithin{equation}{section}

\begin{document}

\title{On Superselection Theory of Quantum Fields \\ in Low Dimensions\footnote{Invited lecture at the
XVIth International Congress on Mathematical Physics. Prague, August 3-8, 2009.}}

\author{Michael M\"uger\footnote{Partially based on joint work with A.~Kitaev and with A.~Davydov,
    D.~Nikshych and V.~Ostrik.} \\
Institute of Mathematics, Astrophysics and Particle Physics. \\
Radboud University, Nijmegen, The Netherlands. E-mail: mueger@math.ru.nl
}

\maketitle

\begin{abstract}
We discuss finite local extensions of quantum field theories in low space time dimensions in
connection with categorical structures and the question of modular invariants in conformal field
theory, also touching upon purely mathematical ramifications.
\end{abstract}

%\keywords{Style file; \LaTeX; Proceedings; World Scientific Publishing.}

%\bodymatter

\section{Introduction: Superselection theory in Local Quantum Theory}
The aim of this contribution is to briefly review some aspects of the `quantum
symmetry' question in low dimensional (typically conformal) quantum field theory (QFT), from the
point of view of algebraic 
quantum field theory. The latter (also called `local quantum physics') arguably is the
most successful approach to the axiomatic study of QFT in Minkowski space (or on more general
pseudo-Riemannian manifolds). Due to space constraints we must refer to \cite{haag} or \cite{halv} 
for typical lists of axioms and recall only that a QFT is a map $\2O\mapsto\2A(\2O)$ assigning to every
decent spacetime region $\2O$ an algebra of operators $\2A(\2O)$ satisfying requirements like
spacelike commutativity, covariance w.r.t.\ the group of spacetime symmetries, irreducibility, etc.
Under these assumptions, augmented by Haag duality, it was shown in \cite{dhr3} that the category
DHR$(\2A)$ of `localized representations' (with finite statistical dimension) of a QFT $\2A$ in
Minkowski space of dimension $\ge 2+1$ is a 
rigid semisimple symmetric tensor $*$-category with simple unit. 
%(A priori, this category is realized by endomorphisms of a $C^*$-algebra associated with $\2A$, but
%not by vector spaces and linear maps between them.) 
Later is was proven \cite{dr1} that every such category is the representation category of a unique
compact group $G$. Furthermore \cite{dr2}, there is a (graded local) QFT $\2F$
with the following properties: (1) It is acted upon by $G$ and $\2F^G\cong\2A$. (2) The category
DHR$(\2F)$ is trivial \cite{cdr}. (This must be modified suitably when $\2A$ has fermionic
representations). (3) Upon restriction to $\2F^G\cong\2A$, the representation space of $\2F$
decomposes into a direct sum of irreps, which are precisely the localized representations of
$\2A$. (4) Every local extension $\2B\supset\2A$ is isomorphic to $\2F^H$ for some closed subgroup
$H\subset G$. As argued in \cite[Section 10.6]{halv}, the above amounts to a very satisfactory
Galois theory for QFTs. \cite{halv} also provides an alternative and more transparent
approach to the construction of $\2F$, based on unpublished work by Roberts and Deligne's
independent approach \cite{del} to proving $\2C\simeq\mathrm{Rep}\,G$ for symmetric categories, cf.\
also \cite{mue16}.
% for a $*$-category version. 

%%%%%%%%%%%%%%%%%%%%%%%%%%%%%%%

\section{What changes in low dimensional spacetimes?}
When one considers QFTs in Minkowski space of dimension $1+1$ or just on the light ray $\7R$, the
category DHR$(\2A)$ can only be proven to be braided \cite[I]{frs}. In \cite{klm} this was strengthened
considerably, in that it was shown that DHR$(\2A)$ is maximally non-symmetric, to wit a modular
category in the sense of Turaev, for the important class of `completely rational' models (a rigorous
axiomatization of rational chiral conformal QFTs). In this situation it is clear that DHR$(\2A)$
cannot be equivalent to a representation category of a group. Another advantage of the alternative
construction of the field net \cite{halv} is that it also applies in low dimensions {\it provided} a
fiber functor from DHR$(\2A)$ to the category of Hilbert spaces is given: Roberts' construction of
the field net as described in \cite{halv} goes through essentially unchanged, and appealing to a
version of Woronowicz's Tannakian theorem \cite{MRT} one obtains a discrete algebraic quantum group
acting on $\2F$ and an R-matrix describing the spacelike commutation relations of $\2F$ (which will
be far from local), cf.\ \cite{mue_ext}.

However, in general a fiber functor for DHR$(\2A)$ will not exist (as can be proven e.g.\ when the
category is finite and contains objects of non-integer dimension). At least in the case of a finite
category, one can always get around this problem obtaining \cite{ostrik} a weak Hopf algebra,
acting on the reduced field bundle of \cite[II]{frs},
but it is not clear that this is useful: The weak Hopf algebra is not uniquely determined by
DHR$(\2A)$ so that it does not have an intrinsic physical meaning. (Furthermore, the inclusion
$\2A\subset\2F$ has the undesirable property of being reducible.) For many purposes it seems better
to resign oneself to considering  the category DHR$(\2A)$ itself as the fundamental structure. The
rest of this note is an attempt to convince the reader of the feasibility, even elegance, of such an
approach. Cf.\ \cite{mue_ext} for details.

%%%%%%%%%%%%%%%%%%%%%%%%%%%%

\section{Categorical analysis of local extensions}
The first question we address is the classification of extensions of a QFT, in particular local ones
and their representation categories. By an extension $\2B\supset\2A$ of a QFT we mean an inclusion
preserving assignment $\2O\mapsto\2B(\2O)\supset\2A(\2O)$ satisfying covariance and irreducibility,
but not necessarily locality. If $\2B$ is local we call it a local extension. In \cite{lre} it was
shown that there is a bijection between (unitary equivalence classes of) extensions $\2B\supset\2A$
that are finite (i.e.\ $[\2B(\2O):\2A(\2O)]$ is independent of $\2O$ and finite) and (equivalence
classes of) Q-systems in the category DHR$(\2A)$. (A Q-system in a $*$-category is essentially a
monoid/algebra $(\Gamma,m,\eta)$ such that $(\Gamma,m,\eta,m^*,\eta^*)$ is a Frobenius algebra.)
Furthermore, $\2B$ is local iff $(\Gamma,m,\eta)$ is commutative, i.e.\ $m\circ c(\Gamma,\Gamma)=m$,
where $c$ is the braiding.\footnote{From now on we will simply write `commutative algebra' with the
understanding that  $\dim\mathrm{Hom}(\11,\Gamma)=1$ and that the algebra is `strongly separable symmetric
Frobenius' or `\'etale' \cite{dmno}.} When this is the case, it is natural to ask for a determination of the
category DHR$(\2B)$ in terms of DHR$(\2A)$ and $(\Gamma,m,\eta)$. The answer is given by:

\begin{theorem} \cite{mue_ext} \label{t0}
Let $B\supset\2A$ be the extension of the completely rational theory $\2A$ corresponding to the
commutative algebra $(\Gamma,m,\eta)$ in DHR$(\2A)$. Then there is an equivalence 
\[ \mathrm{DHR}(B)\simeq \Gamma-\mathrm{Mod}_{\mathrm{DHR}(A)}^0 \]
of braided tensor categories. In particular, $\dim(\mathrm{DHR}(B))=\dim(\mathrm{DHR}(A))/d(A)^2$.
\end{theorem}

\begin{remark} 1. Here,  $\Gamma-\mathrm{Mod}_{\mathrm{DHR}(A)}^0$ is the full subcategory of the
category $\Gamma-\mathrm{Mod}_{\mathrm{DHR}(A)}$ of left $\Gamma$-modules consisting of the objects
$(X,\mu)$ satisfying $\mu\circ c(X,\Gamma)\circ c(\Gamma,X)=\mu$. These 
`dyslexic' or `local' modules form a braided tensor category in a canonical way, cf.\ \cite{pareigis}. 
While the proof of Theorem \ref{t0} is too long to be exhibited here, it can be given in a
few pages including the prerequisites. It uses
results of \cite{BE} on $\alpha$-induction to obtain a full and faithful braided tensor functor
$\Gamma-\mathrm{Mod}_{\mathrm{DHR}(A)}^0\rightarrow\mathrm{DHR}(\2B)$ and \cite{klm} together with the
result $\dim\Gamma-\mathrm{Mod}_{\2C}^0=\dim\2C/d(\Gamma)^2$ from \cite{ko} to conclude essential
surjectivity. 

2. The analogous result in the context of vertex operator algebras appears in \cite{ko}, but many
hard VOA technicalities have not yet been worked out. 
%This should be contrasted with the rather easy proof in the operator algebraic framework.

3. The above result also holds in $\ge 2+1$ dimensions where the representation categories are
symmetric and every $\Gamma$-module is local. But in this case, $\mathrm{DHR}(A)$ is equivalent to
$\mathrm{Rep}\,G$ and commutative algebras in $\mathrm{Rep}\,G$ correspond to subgroups
of $H\subset G$, cf.\ \cite{ostrik}. Thus we recover (at least for finite extensions) the Galois
correspondence mentioned in the introduction. 

4. In view of the above result, almost all questions concerning finite local extensions of
QFTs and their representation categories can be reduced to purely categorical, i.e.\ algebraic
considerations. An obvious exception to this is the inverse problem, to wit the question which
categories are realized in some QFT model. 
(The analogous question in the group theoretic situation $d\ge 2+1$ was settled in \cite{DP}.)
%(In the case $d\ge 2+1$ the analogous question was settled by Doplicer and Piacitelli.)
\end{remark}

Recall that in $d\ge 2+1$ dimensions, the extension $\2F$ has trivial representation category. In
low dimensions it is {\it not} true that a local extension with this property always exists, since
in view of Theorem \ref{t0}, we have:

\begin{corollary} \label{c1}
A completely rational theory $\2A$ admits a local extension with trivial
  representation category iff $\2C=\mathrm{DHR}(\2A)$ contains a commutative algebra $\Gamma$ such
  that $\Gamma-\mathrm{Mod}_\2C^0$ is trivial, equivalently $d(\Gamma)^2=\dim\2C$.
\end{corollary}

(Notice that every commutative algebra $\Gamma$ in a modular category $\2C$ satisfies
$d(\Gamma)^2\le\dim\2C$, cf.\ \cite{ko}.) The question when such a commutative algebra exists is
answered by the following remarkable

\begin{theorem} \label{t1}
A commutative algebra $\Gamma$ in a modular category gives rise to a braided equivalence
%\begin{equation} 
$Z(\Gamma-\mathrm{Mod}_\2C)\simeq\2C\boxtimes\widetilde{\Gamma-\mathrm{Mod}^0_\2C}$.
%\label{eq}\end{equation}
Thus if $\Gamma-\mathrm{Mod}_\2C^0$ is trivial, equivalently $d(\Gamma)^2=\dim\2C$, then $\2C\simeq Z(\2D)$,
where $\2D=\Gamma-\mathrm{Mod}_\2C$ is a fusion category. Conversely, if $\2D$ is a fusion category,
then $Z(\2D)$ contains a commutative algebra $\Gamma$ with $d(\Gamma)^2=(\dim\2D)^2=\dim Z(\2D)$. 
\end{theorem}
Here $Z(\cdot)$ is the categorical center of a tensor category due to Drinfeld, Majid, Joyal and
Street which was proven \cite{mue10} to be modular for every fusion category $\2D$
with $\dim\2D\ne 0$. Theorem \ref{t1} was obtained by Kitaev and the author \cite{KM}. The first 
half was discovered independently in \cite{dgno}, and results related to the second half are
obtained in \cite{BV}. Combining Corollary \ref{c1} and Theorem \ref{t1} we have:

\begin{corollary} \label{c2}
A completely rational theory $\2A$ admits a local extension with trivial
  representation category iff the modular category $\mathrm{DHR}(\2A)$ is a Drinfeld center.
\end{corollary}

The $\Rightarrow$ direction actually seems to be more useful. E.g., one can use it to prove:

\begin{corollary}
Let $\2B$ be a completely rational theory with trivial representation category acted upon freely by
a finite group $G$, and let $\2A=\2B^G$ be the `orbifold' theory. Then there is a unique class
$[\omega]\in H^3(G,\7T)$ such that $\mathrm{DHR}(A)\simeq D^\omega(G)-\mathrm{Mod}$, where
$D^\omega(G)$ is the twisted quantum double of $G$ \cite{dpr}.
\end{corollary}

\begin{remark} 
%1.The corollary amounts to a rigorous formulation of (some of) the ideas in \cite{dvvv}.
A more general analysis of orbifold QFTs $\2A=\2B^G$ without the triviality assumption on
$\mathrm{DHR}(\2B)$ was given in \cite{mue15}. However, this did not yield the above corollary since
a certain coherence theorem for braided crossed G-categories was missing. But see \cite{mue_t}.  
\end{remark}

Even when $\2A$ admits no local extension $\2B\supset\2A$ with trivial $\mathrm{DHR}(\2B)$, it
follows easily from the above results that maximal local extensions always exist and that every
local extension embeds into a maximal one. Contrary to the situation for $d\ge 2+1$, it is not true
that there is a unique (up to unitary equivalence) maximal local extension. (E.g., if $\2A$ is
completely rational with DHR$(\2A)\simeq D(G)-\mathrm{Mod}$, then there exist at least two
non-isomorphic commutative algebras $\Gamma,\widehat{\Gamma}$ such that
$d(\Gamma)=d(\widehat{\Gamma})=|G|$, giving rise to local extensions that are not unitarily equivalent.)

However, by a result of \cite{dmno}, cf.\ Section 5, all maximal local extensions have braided equivalent
representation categories. In particular, if a local extension with trivial representation category
exists, then every local extension embeds into one with trivial representation category.

%%%%%%%%%%%%%%%%%%

\section{Connections with Rehren's approach to modular invariants}
It is well known that a modular category $\2C$ gives rise to a finite dimensional representation of
the modular group $SL(2,\7Z)$, acting on the complexified Grothendieck group of $\2C$.
Given two modular categories $\2C_L,\2C_R$, a {\it modular invariant} used to be defined as a
$\7Z_{\ge 0}$-valued matrix $(Z_{ij})$, indexed by the respective sets $I_L,I_R$ of simple objects,
that satisfies $Z_{00}=1$ and intertwines the associated representations of $SL(2,\7Z)$,
i.e.\ $Z\pi_L(g)=\pi_R(g)Z$. This definition turned out to be insufficient, and from now on `modular 
invariant' will mean the following for us:

\begin{definition} A modular invariant for a pair $(\2C_L,\2C_R)$ of modular categories consists of
  a triple $(\Gamma_L,\Gamma_R, E)$, where $\Gamma_L, \Gamma_R$ are commutative algebras in $\2C_L,
  \2C_R$, respectively, and $E$ is a braided monoidal equivalence
$\Gamma_L-\mathrm{Mod}_{\2C_L}^0\rightarrow\Gamma_R-\mathrm{Mod}_{\2C_R}^0$. 
\end{definition}

The question immediately arises whether two given categories $\2C_L,\2C_R$ admit a modular
invariant. In the left-right symmetric case $\2C_L=\2C_R=\2C$ at least one modular  invariant
always exists, to wit the triple $(\11,\11,\mathrm{id}_\2C)$ (giving rise to the diagonal matrix
$Z=\11$). A very ingenious subfactor theoretic approach to finding non-trivial modular invariants in
this left-right symmetric situation was initiated in \cite{BE} and pursued further in \cite{BEK}.
This approach turned out \cite{ostrik,FRS} to be of essentially categorical nature, 
revolving around algebras in $\2C$ that are not necessarily commutative and thus do not have an
interpretation in terms of local extensions. A more physically motivated approach, applicable also
when $\2C_L\not\simeq\2C_R$,  was proposed in \cite{khr5}, where it was shown that, 
given two chiral CQFTs $A_L,A_R$ and considering their product $\2A=A_L\otimes\overline{A_R}$ as a QFT
on $1+1$-dimensional Minkowski space, every finite local extension $\2B\supset\2A$ gives rise to 
a matrix $Z$ satisfying $Z_{00}=1$ and intertwining the $T$-matrices of the modular categories
$\2C_L,\2C_R$. Furthermore, it was 
conjectured that $ZS_L=S_RZ$ iff $DHR(\2B)$ is trivial. Proofs of this were found by the author and
by R. Longo and Y. Kawahigashi (both remained unpublished). In \cite{khr5} it was also shown that,
given such a local extension $\2B\supset A_L\otimes\overline{A_R}$, there are chiral local extensions
$\widehat{A}_L\supset A_L,\widehat{A}_R\supset A_R$ that are maximal w.r.t.\ the property
$A_L\otimes\overline{A_R}\subset\widehat{A}_L\otimes\overline{\widehat{A}_R}\subset\2B$ and have
isomorphic fusion rings. In \cite{kl} it was even shown
that one has an equivalence $E:\mathrm{DHR}(\widehat{A}_L)\rightarrow\mathrm{DHR}(\widehat{A}_R)$ of braided
tensor categories. Recalling the correspondence \cite{lre} between local extensions $B\supset A$ and
commutative algebras in $\mathrm{DHR}(A)$ as well as Theorem \ref{t0}, we conclude that a commutative algebra
$\Gamma\in\mathrm{DHR}(A_L\otimes\overline{A_R})\simeq\mathrm{DHR}(A_L)\boxtimes\widetilde{\mathrm{DHR}(A_R)}$
of maximal dimension gives rise to a modular invariant for $\2C_L,\2C_R$. The converse follows from
the fact, resulting from \cite{lre} and \cite{klm}, that $A\otimes\overline{A}$ admits a local extension
with trivial representation category. At this point, two questions arise: 
\begin{enumerate}
\item What is the significance of the above reasoning for the construction of $d=2$ CQFTs? Perhaps
the requirement that $\2B$ have trivial representation category amounts to the absence of an
obstruction to existence of a `Wick rotated' Euclidean CQFT that can be defined on arbitrary Riemann
surfaces? (The fact that the DHR category has a cohomological interpretation in terms of Roberts'
local cohomology may be taken as further support for this speculation.)
\item Do the above results hold irrespective of whether the categories $\2C_{L/R}$ arise from chiral
CFTs $A_{L/R}$? The answer is yes:
\end{enumerate}

\begin{theorem} \label{t3}
Given two modular categories $\2C_L,\2C_R$, a modular invariant $(\Gamma_L,\Gamma_R, E)$ gives rise
to a commutative algebra $\Gamma\in\2C_L\boxtimes\widetilde{\2C}_R$ of maximal dimension
(i.e.\ $d(\Gamma)=(\dim\2C_L\cdot\dim\2C_R)^{1/2}$). Conversely, every such algebra arises from a
modular invariant \cite{dmno}, which in fact is unique \cite{dm}.
\end{theorem}

(In the case $\2C_L=\2C_R$ this was proven in \cite{KR}.)
The proof of the first half requires little more than ideas already contained in \cite{lre} and \cite{ko}.
%As to the proof: Given a modular invariant
%$(\Gamma_L,\Gamma_R, E)$, there exists a commutative algebra of 
%maximal dimension in $\Gamma_L-\mathrm{Mod}^0_{\2C_L}\boxtimes\widetilde{\Gamma_R-\mathrm{Mod}^0_{\2C_R}}$, and a
%change-of-base argument gives a maximal commutative algebra in
%$\2C_L\boxtimes\widetilde{\2C_R}$. 
The converse implication follows from facts outlined in the final section.

\section{On the classification of modular categories}
The circle of ideas of the preceding section has important pure mathematical ramifications, cf.\ \cite{dmno}.
Namely, given modular categories $\2C_L,\2C_R$, it is not hard to see that existence of a  fusion
category $\2D$ such that 
$\2C_L\boxtimes\widetilde{\2C_R}\simeq Z(\2D)$ is equivalent to existence of fusion categories
$\2E_1,\2E_2$ such that $\2C_L\boxtimes Z(\2E_1)\simeq\2C_R\boxtimes Z(\2E_2)$. In this case we write
$\2C_L\sim\2C_R$, and one easily sees that $\sim$ is an equivalence relation, 
called Witt equivalence in analogy with the classical theory of quadratic forms. The set $W_M$ of
Witt equivalence classes of modular categories acquires an abelian monoid structure  
by $[\2C_1]\cdot[\2C_2]:=[\2C_1\boxtimes\2C_2]$ and $\11_{W_M}=[\mathrm{Vect}]$.
In view of $\2C\boxtimes\widetilde{\2C}\simeq Z(\2C)$ (by \cite{mue10}) and $Z(\2C)\sim\11$ (by
definition), $W_M$ is a group with inverse operation $[\2C]^{-1}=[\widetilde{\2C}]$.

In view of Theorem \ref{t1}, if $\Gamma$ is a commutative algebra in $\2C$ then
$[\2C]=[\Gamma-\mathrm{Mod}^0_\2C]$. (Thus if $A\subset B$ are completely rational CFTs then 
$[\mathrm{DHR}(A)]=[\mathrm{DHR}(B)]$.) Calling a modular category {\it completely anisotropic} if
every commutative algebra in it is isomorphic to $\11$, one can prove that Witt-equivalent completely
anisotropic categories are actually braided equivalent. (In other words: every Witt equivalence class
contains one completely anisotropic category up to braided equivalence). This is the reason for the
uniqueness of the representation category of maximal local extensions and of the converse statement
in Theorem \ref{t3}: By Theorem \ref{t1}, existence of a maximal commutative algebra in
$\2C_L\boxtimes\widetilde{\2C_R}$ implies $\2C_L\boxtimes\widetilde{\2C_R}\simeq Z(\2D)$ and thus
$[\2C_L]=[\2C_R]$. Choosing commutative algebras $\Gamma_{L/R}\in\2C_{L/R}$ such that the respective
local module categories are completely anisotropic, we have
$[\Gamma_L-\mathrm{Mod}^0_{\2C_L}]=[\2C_L]=[\2C_R]=[\Gamma_R-\mathrm{Mod}^0_{\2C_R}]$ and thus 
a braided equivalence $\Gamma_L-\mathrm{Mod}^0_{\2C_L}\simeq\Gamma_R-\mathrm{Mod}^0_{\2C_R}$.

Finally, the Witt group holds great promise for the
classification of modular categories. The point is that $Z(\2C)$ is modular for every fusion
category, of which there are far too many to hope for a classification. (The fact that inequivalent
fusion categories can have equivalent centers \cite{mue10} does not help much.)
Passing to the Witt group not only kills those `trivial' modular categories but has the nice
effect of yielding an abelian group $W_M$. Many generators (from quantum groups at roots of unity) and
relations (from conformal extensions and cosets) for $W_M$ are already known, and one may hope that
$W_M$ can be determined completely. This would seem to be a rigorous implementation to the idea from
CQFT folkore that ``all modular categories'' arise from the chiral WZW models via local extensions
and coset constructions.

\bibliographystyle{ws-procs975x65}

\end{document}